\documentclass[preprint,showpacs,preprintnumbers,amsmath,amssymb]{revtex4}

\usepackage{graphicx}
\usepackage{dcolumn}
\usepackage{bm}

\begin{document}

\preprint{APS/123-QED}

\title{Analysis of price diffusion in financial markets using PUCK model}

\author{Takayuki Mizuno$^{a}$, Hideki Takayasu$^{b}$, and Misako Takayasu$^{a}$}%
\address{$^{a}$Department of Computational Intelligence {\&} Systems Science, 
Interdisciplinary Graduate School of Science {\&} Engineering, 
Tokyo Institute of Technology, 4259-G3-52 Nagatsuta-cho, Midori-ku, Yokohama 226-8502, Japan \\
$^{b}$Sony Computer Science Laboratories Inc., 3-14-13 Higashigotanda, Shinagawa-ku, Tokyo 141-0022, Japan}

\date{\today}

\begin{abstract}
Based on the new type of random walk process called the Potentials of 
Unbalanced Complex Kinetics (PUCK) model, we theoretically show that the 
price diffusion in large scales is amplified $2(2 + b)^{ - 1}$ times, where 
$b$ is the coefficient of quadratic term of the potential. In short time 
scales the price diffusion depends on the size $M$ of the super moving 
average. Both numerical simulations and real data analysis of Yen-Dollar 
rates are consistent with theoretical analysis.
\end{abstract}

\maketitle

\section{Introduction}

 Crashes and uncontrollable hikes can often occur in financial markets. Such 
changes of the prices confuse the market and damage the economy because they 
start abruptly in many cases. Therefore, techniques to measure the 
probabilistic risk of sudden change in the prices have been studied using 
tick-by-tick data [1]. Recently, it was empirically found that change of 
prices can be approximated by the Fokker-Planck equation and the new type of 
random walk in a potential field [2,3,4,5]. The potential field is 
approximated by a quadratic function with its center given by the moving 
average of past market prices. This random walk model is called the 
Potentials of Unbalanced Complex Kinetics (PUCK) model in which the 
potential slowly changes in the market [3,4]. In this paper, we focus on the 
diffusion properties of this random walk process and calculate the diffusion 
coefficient which is helpful for estimating the market risk.

 We first review an empirical derivation of the PUCK model. We next show 
that the statistically steady condition of price fluctuations depends on the 
potential field, and clarify relationships between the price diffusion and 
the potential field. We finally demonstrate that the price diffusion in 
short time scales depends on the size of moving average, however, large 
scale diffusion properties are independent of the moving average. In the 
paper, we used all the Bid record (about 20 million ticks) of the exchange 
rates for Yen/Dollar that were traded by the term from 1989 to 2002 to find 
the firm statistical laws.

\section{Empirical derivation of PUCK model}

 Prices in financial markets always have violent fluctuation in a short time 
scale. We first eliminate the uncorrelated noise $\eta (t)$ from the 
price $P(t)$ in order to reduce the statistical error. We next investigate 
the dynamics of the price.

We can perform this noise elimination process by introducing an optimum 
moving average $\overline {P(t)} $:

\begin{equation}
\label{eq1}
P(t) = \overline {P(t)} + \eta (t),
\end{equation}

\begin{equation}
\label{eq2}
\overline {P(t)} = \sum\nolimits_{k = 1}^K {w_k \cdot P(t - k)} ,
\end{equation}

\noindent
where $P(t)$ is a price, $\eta (t)$ is an uncorrelated noise and $w_k $ 
gives the weight factors where the time is measured by ticks. The weight 
factors are calculated by using Yule-Walker equation [6,7,8]. In a case of 
Yen-Dollar rate, the weight factor $w_k $ follows an exponential function 
whose characteristic decay time is about 30 seconds [6].

 We investigate a behavior of the optimum moving average $\overline {P(t)} $ 
obtained by eliminating the uncorrelated noise $\eta (t)$ from the price 
$P(t)$. We introduce a super moving average $\overline {P_M (t)} $ defined 
by

\begin{equation}
\label{eq3}
\overline {P_M (t)} = \frac{1}{M}\sum\nolimits_{\tau = 0}^{M - 1} {\overline 
{P(t - \tau )} } .
\end{equation}

\noindent
In financial markets, it is found that the following relationship between 
$\overline {P(t)} $ and $\overline {P_M (t)} $ holds for a certain range of 
$M$[2],

\begin{equation}
\label{eq4}
\overline {P(t + 1)} - \overline {P(t)} = - \frac{1}{2} \cdot \frac{b(t)}{M 
- 1} \cdot \frac{d}{d\overline P }\left( {\overline {P(t)} - \overline {P_M 
(t)} } \right)^2 + f(t),
\end{equation}

\noindent
where the mean of noise $f(t)$ is zero. The Eq.(\ref{eq4}) means that the price 
change can be approximated by a random walk in a quadratic potential field 
whose center is given by the moving average of past prices. It is known that 
the potential coefficient $b(t)$ has a long autocorrelation [3,4].

\section{Statistically steady condition of price fluctuations in the PUCK 
model}

 We focus on Eq.(\ref{eq4}) with the case of a constant $b$ because the coefficient 
$b(t)$ is known to change slowly in financial markets. Eq.(\ref{eq4}) is transformed 
as follows,

\begin{equation}
\label{eq5}
\overline {P(t + 1)} - \overline {P(t)} = - \frac{b}{2}\left( {\frac{2}{M(M 
- 1)}\sum\limits_{k = 1}^{M - 1} {(M - k)\left( {\overline {P(t - k + 1)} - 
\overline {P(t - k)} } \right)} } \right) + f(t).
\end{equation}

\noindent
This is a type of AR process for price difference when $b$ is a constant. We 
can estimate the conditions of $b$ to make the AR process being 
statistically steady. Eq.(\ref{eq5}) is transformed by the following determinant,

\begin{eqnarray}
\label{eq6}
 X_t &=& AX_{t - 1} + F_t \nonumber \\ 
 &=& \left( A \right)^tX_0 + \left( {F_t + AF_{t - 1} + \cdots + \left( A 
\right)^{t - 1}F_1 } \right),
\end{eqnarray}

\noindent
where

\begin{equation}
\label{eq7}
{\begin{array}{*{20}c}
 {X_t = \left( {{\begin{array}{*{20}c}
 {\overline {P(t + 1)} - \overline {P(t)} } \hfill \\
 {\overline {P(t)} - \overline {P(t - 1)} } \hfill \\
 {\overline {P(t - 1)} - \overline {P(t - 2)} } \hfill \\
 \vdots \hfill \\
 {\overline {P(t - M + 3)} - \overline {P(t - M + 2)} } \hfill \\
\end{array} }} \right),} \hfill & {A = \left( {{\begin{array}{*{20}c}
 {\alpha _1 } \hfill & {\alpha _2 } \hfill & \cdots \hfill & {\alpha _{M - 
2} } \hfill & {\alpha _{M - 1} } \hfill \\
 1 \hfill & 0 \hfill & \cdots \hfill & 0 \hfill & 0 \hfill \\
 0 \hfill & 1 \hfill & \cdots \hfill & 0 \hfill & 0 \hfill \\
 \vdots \hfill & \vdots \hfill & \ddots \hfill & \vdots \hfill & \vdots 
\hfill \\
 0 \hfill & 0 \hfill & \cdots \hfill & 1 \hfill & 0 \hfill \\
\end{array} }} \right),} \hfill \\
 {F_t = \left( {{\begin{array}{*{20}c}
 {f(t)} \hfill \\
 0 \hfill \\
 0 \hfill \\
 \vdots \hfill \\
 0 \hfill \\
\end{array} }} \right),} \hfill & { \alpha _k = - \frac{b(M - k)}{M(M - 
1)}.} \hfill \\
\end{array} } 
\end{equation}

\noindent
If $\mathop {\lim }\limits_{t \to \infty } \left( A \right)^t = 0$, the time 
series of $\overline {P(t + 1)} - \overline {P(t)} $ is in a weakly steady 
state because $X_t $ becomes independent of the initial value of $X_0 $. 
This condition is fulfilled when the absolute values of all eigenvalues of 
$A$ are less than 1. For example, when $M = 2$ the time series of $\overline 
{P(t + 1)} - \overline {P(t)} $ is statistically steady if the potential 
coefficient satisfies $\left| b \right| < 2$. When $M = 3$ the eigenvalues 
of $A$ are given by

\begin{equation}
\label{eq8}
\lambda _1 = \frac{\alpha _1 + \sqrt {\alpha _1^2 + 4\alpha _2 } }{2},
\quad
\lambda _2 = \frac{\alpha _1 - \sqrt {\alpha _1^2 + 4\alpha _2 } }{2}.
\end{equation}

\noindent
Solving the steady state condition we find that the potential coefficient 
should be in the range of $ - 2 < b < 6$ for $M = 3$. Numerically 
calculating the eigenvalues for $2 \le M \le 20$ as shown in Fig.1 we find 
that the time series of $\overline {P(t + 1)} - \overline {P(t)} $ is 
statistically stationary if the potential coefficient is in the following 
range,

\begin{equation}
\label{eq9}
\left\{ {{\begin{array}{*{20}c}
 { - 2 < b < 2(M - 1)} \hfill \\
 { - 2 < b < 2M} \hfill \\
\end{array} }} \right.
\quad
{\begin{array}{*{20}c}
 {\mbox{when}} \hfill \\
 {\mbox{when}} \hfill \\
\end{array} }
\quad
{\begin{array}{*{20}c}
 M \hfill \\
 {M } \hfill \\
\end{array} }
{\begin{array}{*{20}c}
 {\mbox{is}} \hfill \\
 {\mbox{is} } \hfill \\
\end{array} }
{\begin{array}{*{20}c}
 {\mbox{even}} \hfill \\
 {\mbox{odd}} \hfill \\
\end{array} }
{\begin{array}{*{20}c}
 {\mbox{number}} \hfill \\
 {\mbox{number} } \hfill \\
\end{array} }.
\end{equation}

\noindent
Outside the condition of Eq.(\ref{eq9}), the range of price fluctuations increases 
indefinitely depending on the time $t$.

\section{Diffusion of prices in market potential field}

 As the potential coefficient $b(t)$ has a long autocorrelation, we can 
calculate the future price diffusion using Eq.(\ref{eq4}). This prediction is 
crucial in order to evaluate the risks of market. We clarify statistical 
laws of price diffusion described by Eq.(\ref{eq4}) using both simulations and 
theoretical analysis. 

 By simulating Eq.(\ref{eq4}) for the case of $b(t)$ is a constant, we investigate 
the standard deviation on a time scale $T$ defined by

{\small 
\begin{equation}
\label{eq10}
\sigma _b (T) = \sqrt {\left\langle {\left( {\overline {P(t + T)} - 
\overline {P(t)} } \right)^2} \right\rangle } .
\end{equation}
}

\noindent
In Fig.2 we plot $\sigma _b (T)$ for $M = 4$, $16$, $64$, $256$ when $b = - 
1.5$ and $b = 2$. Here, $f(t)$ is the Gaussian random number whose standard 
deviation is 1. The time scale where the Hurst exponent converges 0.5 
depends on $M$. For example, the time scale is about $10^3$ when $M = 256$, 
while the time scale is around $T = 20$ when $M = 4$

We can estimate the standard deviation $\sigma _b (T)$ in the long time 
scale limit. The variance of change of optimum moving average price 
$\overline {P(t)} $ is given by

{\footnotesize 
\begin{eqnarray}
\label{eq11}
 \left\langle {\left( {\overline {P(t + T)} - \overline {P(t)} } \right)^2} 
\right\rangle &=& - \left( {\frac{b}{2}} \right)^2\left( {\frac{2}{M(M - 1)}} 
\right)^2\left\langle {\left( {\sum\limits_{k = 1}^{M - 1} {(M - k)\left( 
{\overline {P(t - k + T)} - \overline {P(t - k)} } \right)} } \right)^2} 
\right\rangle \nonumber \\ 
 & &- b\frac{2}{M(M - 1)}\left\langle {\sum\limits_{k = 1}^{M - 1} {\left( {(M 
- k)\left( {\overline {P(t + T)} - \overline {P(t)} } \right)\left( 
{\overline {P(t - k + T)} - \overline {P(t - k)} } \right)} \right)} } 
\right\rangle \nonumber \\ 
 & &+ \left\langle {\left( {\sum\limits_{\tau = 0}^{T - 1} {f(t + \tau )} } 
\right)^2} \right\rangle .
\end{eqnarray}
}

\noindent
By introducing a rough approximation

\begin{equation}
\label{eq12}
\left\langle {\left( {\overline {P(t + T)} - \overline {P(t)} } \right) 
\cdot \left( {\overline {P(t + T - k)} - \overline {P(t - k)} } \right)} 
\right\rangle \approx \left\langle {\left( {\overline {P(t + T)} - \overline 
{P(t)} } \right)^2} \right\rangle ,
\end{equation}

\noindent
we have the following simple formulation after some calculation,

\begin{equation}
\label{eq13}
\sigma _b (T) = \left( {\frac{2}{2 + b}} \right) \sigma _{b = 0} (T),
\end{equation}

\noindent
where $T \gg M$, $\sigma _b (T) = $ {\scriptsize $\sqrt {\left\langle {\left( {\overline 
{P(t + T)} - \overline {P(t)} } \right)^2} \right\rangle } $} and $\sigma _{b 
= 0} (T)$ is the standard deviation when $b = 0$. In this long time scales, 
the standard deviation of the price is amplified $2(2 + b)^{ - 1}$ times by 
the potential field. In Fig.3 we show relationships between the potential 
coefficient $b$ and the ratio of standard deviation ${\sigma _b (T)} 
\mathord{\left/ {\vphantom {{\sigma _b (T)} {\sigma _{b = 0} }}} \right. 
\kern-\nulldelimiterspace} {\sigma _{b = 0} }(T)$ when $T = 10^5$ by 
simulating Eq.(\ref{eq4}) for $M = 2$, 4, 8, 16, 32. We can confirm that the price 
diffusion of numerical simulations follows Eq.(\ref{eq13}) independent of $M$ in the 
long time scale. From Eq.(\ref{eq13}) we can theoretically find that in the long 
time scales the price diffusion is independent of $M$, independent of $b$, 
$\sigma _b (T) \propto T^{0.5}$.

\section{Diffusion of Yen-Dollar rates}

From dataset of real market prices, we can estimate the value of $b$, and 
presume the best value of $M$ by comparing the price diffusion of numerical 
simulations to real price diffusion. Fig.4 shows the diffusions of 
Yen-Dollar rates from 3:35 to 8:35 and from 9:25 to 23:25 in 11/9/2001, the 
day of terrorism. The rates were stable till 8:35 and it became quite 
unstable after 9:25. The rates until 8:35 follow a slow diffusion in short 
time scales, namely, the market has an attractive potential $b > 0$. As 
shown in Fig.4, we can recreate the rate diffusion in all time scales by 
simulating Eq.(\ref{eq4}) with $b = 0.9$ and $M = 14$ ticks. We next focus on the 
unstable rates after 9:25 that follow a fast diffusion in short time scales. 
When $b = - 0.8$ and $M = 14$ ticks, the price diffusion of numerical 
simulation is also consistent with the rate diffusion. The price diffusion 
depends on $M$ in the short time scales, although the price diffusion is 
independents of $M$ in the long time scales. Therefore, if $M \ne 14$ ticks, 
there are gaps between the price diffusion of numerical simulations and the 
rate diffusion in the short time scales as shown in Fig.4. In other real 
markets, we can also estimate the best value of $M$ from such 
characteristics of price diffusion.

\section{Discussion}

 We can approximate the change of market prices by the random walk in a 
potential field. The potential field is well approximated by a quadratic 
function with its center given by the moving average of past prices. The 
random walk process is called the PUCK model. By analyzing the model, we 
clarified that the statistically steady condition of price fluctuations 
depends on the potential coefficient $b$, and we also theoretically proved 
that the price diffusion in the long time scales is amplified $2(2 + b)^{ - 
1}$ times, independent of the size $M$ of super moving average. In short 
time scales the price diffusion depends on $M$. We can estimate the best 
value of $M$ in real financial markets by observing this dependence. We 
recreated the diffusion of Yen-Dollar rates in all time scales by the PUCK 
model. The potential coefficient $b$ is helpful to measure the probabilistic 
risk of sudden change in the prices. We may be able to build better 
financial options that offset the risk by applying the price diffusion of 
the PUCK model.

\begin{acknowledgments}
This work is partly supported by Research Fellowships of the Japan Society 
for the Promotion of Science for Young Scientists (T.M.). The authors 
appreciate H. Moriya of Oxford Financial Education Co Ltd. for providing the 
tick data.
\end{acknowledgments}

\newpage

\begin{figure}
\centerline{\includegraphics[width=3.3525in,height=2.7975in]{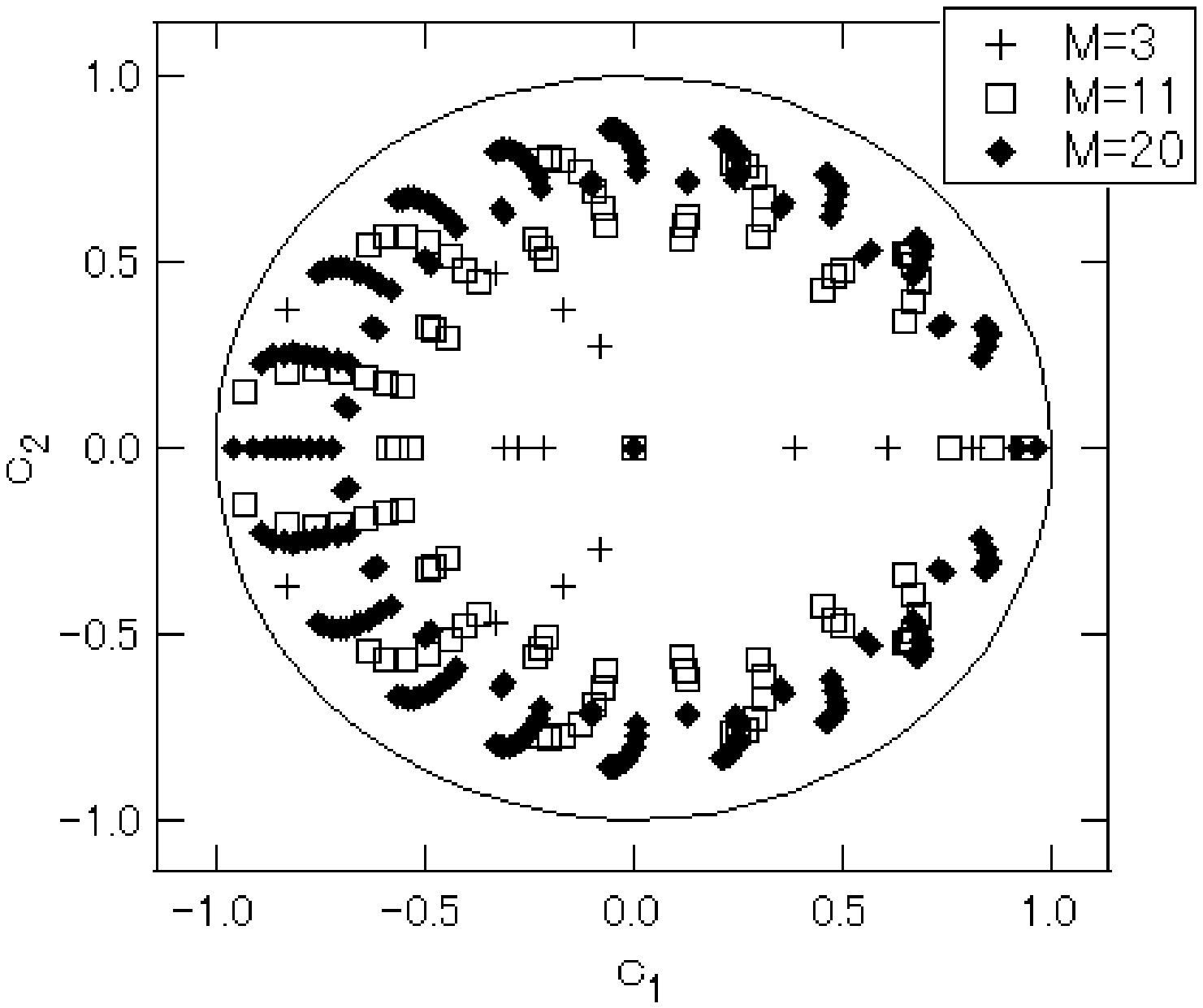}}
\caption{The eigenvalues $\lambda $ of matrix $A$. The $c_1 $ and $c_2 $ are 
real and imaginary parts of $\lambda $. The pluses are $\lambda $ for $M = 3$ 
when $ - 2 < b < 6$. The squares show $\lambda $ for $M = 11$ when $ - 2 < b 
< 22$. The diamonds express $\lambda $ for $M = 20$ when $ - 2 < b < 38$. 
The circle indicates $\left| \lambda \right| = 1$.}
\end{figure}

\begin{figure}
\centerline{\includegraphics[width=6.34in,height=2.83in]{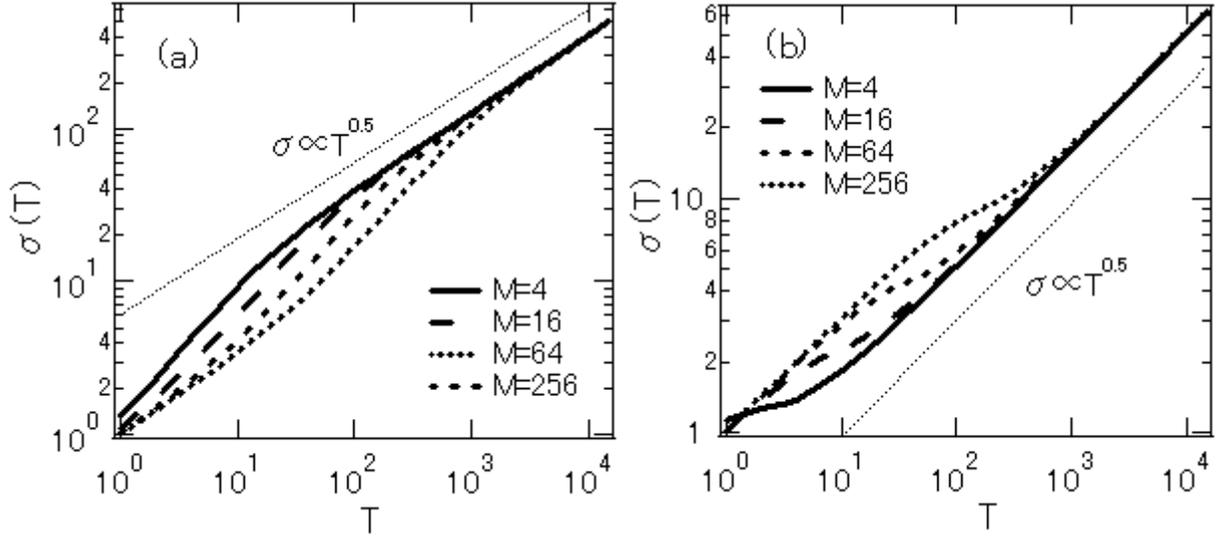}}
\caption{Standard deviation $\sigma (T)$ of price changes in the time scale 
$T$. The heavy lines show price diffusions of numerical simulation for $M = 
4$(top), 16(second), 64(third), 256(bottom). Here, $b = - 1.5$ in (a) 
and $b = 2$ in (b). The standard deviation of $f(t)$ is 1. The guideline 
indicates $\sigma (T) \propto T^{0.5}$.}
\end{figure}

\begin{figure}
\centerline{\includegraphics[width=4.22in,height=1.89in]{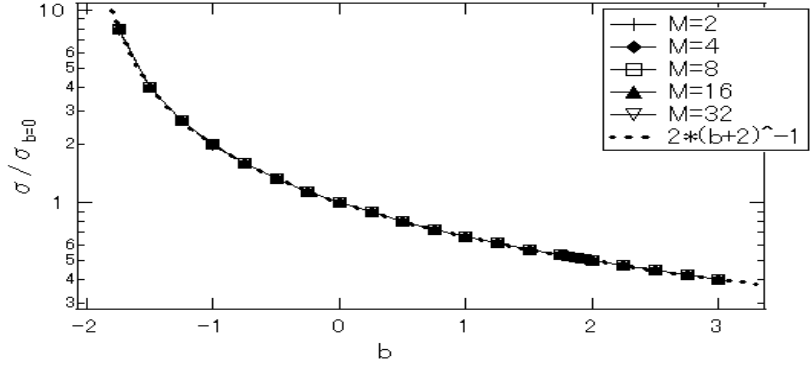}}
\caption{Relationship between potential coefficient $b$ and standard deviation 
ratio ${\sigma _b (T)}$/${\sigma _{b = 0} (T)}$ when 
the time scale $T = 10^5$. Theory (dashed line) and numerical simulations 
for $M = 2$(plus), 4(diamond), 8(square), 16(black triangle), 32(white 
triangle).[4]}
\end{figure}

\begin{figure}
\centerline{\includegraphics[width=3.36in,height=3.39in]{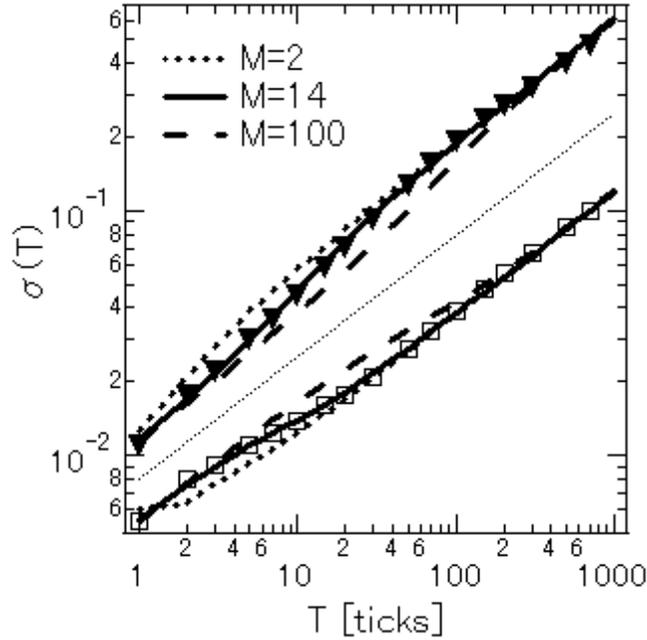}}
\caption{Diffusion of Yen-Dollar rates. The squares and triangles show the 
diffusion from 3:35 to 8:35 and from 9:25 to 23:25 in 11/9/2001. The top 3 
lines are the numerical simulation of Eq.(\ref{eq4}) for $M = 2$, 14, 100 ticks when 
$b = - 0.8$. The standard deviation of $f(t)$ is 0.0115 yen/dollar. The 
bottom 3 lines show the numerical simulation for $M = 2$, 14, 100 ticks when 
$b = 0.9$. The standard deviation of $f(t)$ is 0.0054 yen/dollar. The straight 
line expresses the normal diffusion with the slope 0.5.}
\end{figure}

\end{document}